\documentstyle[11pt,aaspp4]{article}
\lefthead{Gay \& Lambert}
\righthead{Magnesium Isotopic Abundances}

\begin{document}

\title{The Isotopic Abundances of Magnesium in Stars}

\author{Pamela L. Gay and David L. Lambert}
\affil{Department of Astronomy, University of Texas at Austin, Austin TX
78712-1083}

\begin{abstract}
Isotopic abundance ratios $^{24}$Mg:$^{25}$Mg:$^{26}$Mg are derived for 20
stars from high-resolution spectra of the MgH A-X 0-0 band at 5140\AA. With
the exception of the weak G-band giant HR\,1299, the stars are dwarfs that sample
the metallicity range  -1.8 $<$ [Fe/H] $<$0.0. The  abundance of $^{25}$Mg
and $^{26}$Mg relative to the dominant isotope $^{24}$Mg decreases with
decreasing [Fe/H] in fair accord with  predictions from a recent model
of galactic chemical evolution in which the Mg isotopes are
synthesised by massive stars. Several stars appear especially enriched in
the heavier Mg  isotopes suggesting contamination by material from the
envelopes of intermediate-mass AGB stars.

\end{abstract}

\newpage
\twocolumn
\section{Introduction}

Observational investigations of the chemical evolution of the
Galaxy involve the determination and interpretation
 of elemental abundances for stellar samples.  Many studies draw on the
local stellar population  which is a mix of disk and halo stars. 
Results of the abundance analyses are commonly expressed as the variation
of the elemental abundance with respect to the iron abundance, i.e.,
the run of [el/Fe] versus [Fe/H] where, as usual,
[X] = log$_{10}$(X)$_{star}$ - log$_{10}$(X)$_{\odot}$. In this paper,
we present measurements of the relative abundances of the 
stable isotopes of magnesium of which the major isotope is the
even-even nucleus $^{24}$Mg and the minor isotopes are the neutron-rich
$^{25}$Mg and $^{26}$Mg. Our measurements of stellar isotopic
abundances are made from high-resolution spectra of a portion of the
MgH A$^2\Pi$ - X$^2\Sigma^+$ $\Delta v$ = 0 band near 5140\AA.
Approximately 20 stars with iron abundance from [Fe/H] $\simeq$ 0 to
$\simeq -1.8$ were analysed for their Mg isotopic ratios which
are compared to  the predicted evolution
of these  ratios.

Synthesis of magnesium occurs primarily in  the carbon and neon burning
shells of massive stars prior to their deaths
as Type II supernovae (Arnett \& Thielemann 1985). In the carbon shell whose
composition is affected by prior He-burning, $^{25}$Mg and $^{26}$Mg are
abundant because He-burning activated the neutron source
 $^{22}$Ne($\alpha,n)^{25}$Mg and the  neutron capture reaction
 $^{25}$Mg($n,\gamma)^{26}$Mg converted some $^{25}$Mg to $^{26}$Mg.
 The yield relative to $^{24}$Mg
 of the neutron-rich isotopes -
$^{25}$Mg and $^{26}$Mg - is predicted to increase with the 
 initial metallicity of the massive star, and, more specifically, with
 the neutron excess. In the case of the unburnt carbon layers, a key factor
is the 
 $^{22}$Ne abundance which
is  determined by a star's initial abundance of C, N, and O. 
Hydrogen burning by the CNO-cycles ensures that
the initial combined abundances of the C, N, and O are converted primarily
to $^{14}$N which during He-burning is converted to $^{22}$Ne through
to $\alpha$-captures.
Models of Population I and II 25$M_{\odot}$ stars evolved
through hydrostatic burning and through the supernova explosion confirm
the simple expectations 
 (Woosley \& Weaver 1982).

Observational evidence
 of a decline in metal-poor stars  of the  abundance
of $^{25}$Mg and $^{26}$Mg relative to $^{24}$Mg
was provided first by Tomkin \& Lambert (1980) from  an
analysis of high-resolution spectra of MgH lines in the subdwarf
Gmb\,1830 (HD\,103095). The observed  abundances of $^{25}$Mg and
$^{26}$Mg were in fair agreement with Woosley \& Weaver's calculations.
 Additional studies of stars spanning the
range [Fe/H] $\simeq$ -1 to $\simeq$ +0.5 have been reported -
see early work by Boesgaard (1968) and Bell \& Branch (1970) and
more recent work with spectra recorded with solid state detectors by
Barbuy (1985, 1987), Barbuy, Spite, \& Spite (1987), Lambert \&
McWilliam (1986), and
McWilliam \& Lambert (1988).

Recently, Timmes, Woosley, \& Weaver (1995)  have  taken an extensive
grid of predicted yields of Mg isotopes 
(and many other nuclides) from massive stars (Woosley \& Weaver 1995), and built a
model of chemical evolution of the solar neighborhood that predicts the
relative abundance of many elements including the Mg isotopes: 
 the predicted isotopic abundance ratios
$^{25}$Mg/$^{24}$Mg and $^{26}$Mg/$^{24}$Mg decrease sharply with decreasing
metallicity [Fe/H].
The principal goal of our new investigation was
to provide accurate determinations of the isotopic Mg abundances for
a  sample of stars with which to test  Timmes et al.'s
predictions.

\section{Observations}

The stars listed in Table 1 were observed at McDonald Observatory with
the
2.7m Harlan J. Smith reflector and its coud\'{e} spectrograph. Two cameras
were used for this project. With the 6-foot camera, a single order of
an echelle grating was isolated with an interference filter: a 10\AA\
interval around 5135\AA\ was recorded on a Tektronix CCD
at a resolving power of approximately
150,000.  Observations were also
made with the {\it 2dcoud\'{e}} camera (Tull et al. 1995)  where multiple
orders of an echelle are separated by prisms and recorded on a
Tektronix 2048 $\times$ 2048 CCD. 
At the order used for our observations, spectral coverage
is not continuous. A majority of the observations were made with the
interval 5125 to 5145\AA\ in the central order placed on the CCD.
Figure 1 shows a large part of this order for the slightly metal-poor star
HD\,55458 (G88-14). Principal lines of $^{24}$MgH are
identified below the spectrum.
The resolving power was about 160,000 for most observations but 200,000 for
some. When
necessary, multiple exposures of about 30 minutes were co-added to achieve the
desired high signal-to-noise
ratio. All observations were reduced in the standard way using IRAF
procedures.

MgH lines are spread across the observed regions but few lines are
suitable for analysis of the isotopic abundances owing to blending
with identified and unidentified lines. McWilliam \& Lambert (1988)
recommended three MgH lines in the observed window
 for use in determining the isotopic
abundances. These are shown in  Figure 2
  which gives  expanded plots of the
 spectrum of G\,88-14. The recommended feature at
5134.6\AA\ which is a blend of the Q$_1$(23) and R$_2$(11) lines from the
0-0 band; McWilliam and Lambert give accurate wavelengths for all
isotopic components for this and their other recommended features. The
line is obviously asymmetric with a trailing red wing due to the
contributions of the less abundant species $^{25}$MgH and $^{26}$MgH.
This line is flanked by slightly weaker MgH lines that also feature
a trailing red wing. Comparison with synthetic spectra shows that, although
 the
red wings of these lines appear to be carbon copies of the 5134.6\AA\ line,
 other lines contribute such that the isotopic abundances cannot
be reliably extracted from them (Tomkin \& Lambert  1980; McWilliam \&
Lambert 1988). The other recommended MgH features are shown too in
Figure 2. The line at 5138.7\AA,  a blend of the $^{24}$MgH lines
0-0 Q$_1$(22) and 1-1 Q$_2$(14), is in the wing of strong atomic lines. The
third feature at 5140.2\AA\ is also an unresolved blend of two $^{24}$MgH
lines (0-0 R$_1$(10) and 1-1 R$_2$(4)). At our resolution, the $^{26}$MgH
pair of lines is effectively resolved from the $^{24}$MgH and $^{25}$MgH
lines.   
Our spectra are of a  resolving power greater
by a factor of 2 to 3
than  previously used for Mg isotopic analyses.

\section{Analysis}

Since the method of analysis follows closely that described by
McWilliam \& Lambert (1988) and earlier by Tomkin \& Lambert (1980),
a brief description will suffice.
Synthetic spectra are generated and fitted to the observed spectrum.
Three MgH features (Figure 2)
 are used to extract the isotopic
abundances -- see McWilliam \& Lambert  (their Table 2) for the
description of the features between 5134 and 5140\AA\ and their Table 3 for
laboratory measurements of these and other MgH lines. All known MgH
lines were included in  the synthetic spectra, as well as many atomic
lines.

Model atmospheres were computed using the program ATLAS9 (Kurucz 1993).
Defining parameters T$_{\rm eff}$, log g, and [Fe/H] were set as
follows: [Fe/H] is provided from the Str\"{o}mgren indices ({\it b - y}),
$m_1$,
and $c_1$ (Schuster \& Nissen (1989, eqn. 3);  T$_{\rm
eff}$ is  estimated from Str\"{o}mgren indices ({\it b - y}) and $c_1$,
and [Fe/H] using
a relation from Alonso et al. (1996, equation 9); log g is obtained from the
Hipparcos parallax using the recipe described by Nissen, H$\o$g, \& Schuster
(1997). Adopted parameters are summarized in Table 1. We also give there
[Fe/H] determined from spectroscopy by Tomkin \& Lambert (1999), or
 as found  from a search
 of references provided by
SIMBAD.
Adopting the spectroscopic [Fe/H] in the computation of the model atmosphere
would have no impact on the determination of the isotopic Mg abundances. The
exercise was carried out primarily to ensure that, in a comparison with
theoretical predictions (Sec. 4), the isotopic Mg abundances were paired with 
a reliable estimate of [Fe/H].

 In general, the 
photometric and spectroscopic estimates of [Fe/H] are in good agreement.
In discussing the relation between isotopic Mg abundances and
[Fe/H], we have given precedence to the spectroscopic [Fe/H] which is often
based on several consistent determinations even when a single
reference is cited in Table\,1. The accuracy of the [Fe/H] estimate
is probably $\pm$ 0.1 dex or slightly better. There is one case for
which photometric and spectroscopic estimates differ greatly: HD\,108564
for which [Fe/H] = -0.52 from photometry and -1.18  from spectroscopy.
This star is discussed later as possibly `peculiar' from the point of view of 
the isotopic Mg abundances. A second star deserves comment: G87-27 was
given a metallicity of -1.45 by Carney et al. (1994) but our syntheses of
atomic lines gives [Fe/H] = -0.42, a value in good agreement with the
photometric determination. We adopt our spectroscopic estimate but
reinvestigation would be desirable as our spectral window contains very
few unblended weak atomic lines.

The synthetic spectrum  program, MOOG, based on the original version
by Sneden (1973) was used.  A
microturbulence of 1 km s$^{-1}$ was assumed for all the stars, and the
macroturbulence in the range of 1 -- 3.5 km s$^{-1}$
was adjusted to match observed line widths.  The instrumental profile was
taken from thorium lines in the spectrum of a Th-Ar hollow
cathode lamp. The C$_2$ molecule's
Swan system contributes many lines. Following Tomkin \& Lambert (1980),
the strength of a C$_2$ line was fixed using a close triplet of lines
near 5135.6\AA\   (Figure 2).
 In the coolest stars, an unidentified line contributes
to the stellar feature at this wavelength but the C$_2$ contribution
remains measureable.
Lines of the following elements were also included: C, Mg, Sc, Ti, Cr, Fe,
Co, Ni and Y lines but in most cases made no significant contribution to
the chosen MgH features.

For each star, an initial synthetic spectrum was computed
with the isotopic abundances set to  ratios of
$^{24}$Mg:$^{25}$Mg:$^{26}$Mg=80:10:10, a mix essentially equal to the
terrestrial benchmark ratios of 78.99:10.00:11.01 (de Bi\`{e}vre
\& Barnes 1985). We refer to these ratios as the solar ratios.
The Mg abundance was adjusted to fit the $^{24}$MgH
lines of the three `isotopic' features and of other clean MgH lines.
Then, the $^{25}$Mg and $^{26}$Mg abundances were adjusted by trial and
error until the profiles of the three recommended features were
fit satisfactorily. Comparison of the observed and best-fitting synthetic
spectra clearly shows that, although other MgH features clearly indicate the
presence of $^{25}$MgH and $^{26}$MgH lines, features other than the recommended
trio are stronger than predicted indicating either the presence of
unidentified weak blends or quite substantial departures from the predicted
rotational line strengths of the MgH lines. Differences between observed and
best-fitting synthetic spectra are similar from one star to the next.

A sample comparison of observed and synthetic spectra is shown in
Figure 3 for HD\,23439A, a metal-poor star ([Fe/H] = -1.1). The 
adopted best fit is for the ratios $^{24}$Mg:$^{25}$Mg:$^{26}$Mg
 = 78:13:9. The presence of $^{25}$Mg and $^{26}$Mg is umistakeable,
as shown by the very poor fit of the synthetic spectrum (dashed line) computed with
$^{24}$Mg alone (and other non-MgH lines). All MgH lines in the illustrated
interval confirm the presence of $^{25}$MgH and $^{26}$MgH. Equally good
fits are obtained to the three recommended MgH features. The dotted
lines corresponding to ratios of 72:16:12 and 83:10:6,
 which are clearly unacceptable
fits to the observed spectrum, provide an indication of the measurement
uncertainties. Note that MgH lines other than the recommended trio (e.g., lines at
 5134.2\AA, 5135.1\AA, and 5138.3\AA) have
stronger red wings than predicted by the best-fitting synthetic spectrum.

A second comparison (Figure 4) involves the N-rich dwarf HD\,25329
whose MgH lines are stronger than those of HD\,23439A; HD\,25329
is more metal-poor by about 0.7 dex but of a lower temperature. 
Again, the three recommended MgH features give consistent isotopic
ratios. The best fit is obtained with the ratios
$^{24}$Mg:$^{25}$Mg:$^{26}$Mg  = 85:8:8. The dotted lines show 
synthetic spectra corresponding to the ratios 89:6:5 and 81:10:9
 that are considered
not to fit the observed spectrum.

The third selected comparison  (Figure 5) is for 12\,Oph, a star of
solar metallicity. In this case, the recommended features at 5138.7\AA\
and 5140.2\AA\ give very similar results; note how essential it is to fit the
wings of the atomic lines near 5139.3\AA. The third recommended feature at
5134.6\AA\ is obviously contaminated with a blend in its red (and also in
the blue) wing. Based on the two apparently unblended features, the
best fit is for $^{24}$Mg:$^{25}$Mg:$^{26}$Mg = 74:12:13 with the
dotted lines providing unsatisfactory fits with the ratios 69:15:16
and 80:9:10.

In the above examples, syntheses represented by the dotted lines correspond
to isotopic mixtures that clearly do not fit the recommended MgH features.
If the isotopic ratios were the sole adjustable variable, the error in the
ratios would be less than the few per cent error represented by the
dotted lines. An allowance for the observed spectrum's
S/N ratio and a consideration of
 the precision with which the continuum may be set would not increase the
errors beyond those represented by the dotted lines.
Although the predicted strength of a MgH line is sensitive to the adopted
atmospheric parameters, the isotopic ratios are quite insensitive. In those
cases where the $^{24}$MgH lines
are strong, the abundance ratio of
$^{25}$Mg and $^{26}$Mg with respect to $^{24}$Mg is most sensitive to
the adopted microturbulence. Alternative choices for the microturbulence
lead to a change in the $^{24}$Mg abundance needed to fit the cores of the
MgH features but to very little to no change in the red wing to which the
$^{25}$MgH and $^{26}$MgH lines contribute. In the case of HD\,23439A, for
example, a reduction of the microturbulence from 1.0 km s$^{-1}$ to
0.5 km s$^{-1}$ requires an increase in the $^{24}$Mg abundance
by about 0.02 dex which corresponds to a less than 1\% reduction in the
isotopic abundance ratios $^{25}$Mg/$^{24}$Mg and $^{26}$Mg/$^{24}$Mg.
An increase of the microturbulence from 1.0 to 1.5 km s$^{-1}$  results
in a smaller percentage increase of the ratios. The effects are slightly
larger for the stars with the strongest MgH features. In summary,
if the isotopic abundances $^{24}$Mg:$^{25}$Mg:$^{26}$Mg
are expressed as x:y:z where x + y + z = 100, the errors in y and z are
about $\pm$ 2 or slightly better for metal-poor stars and possibly
worse for cool solar-metallicity stars like 12\,Oph. Greater blending
of the $^{25}$MgH feature
with the $^{24}$MgH line  leads to a somewhat  greater uncertainty
for the ratio y/x. If an isolated MgH were available in the
spectra, comparison of the  red isotope-contaminated
wing with the blue wing would provide the combined ratio 
($^{25}$Mg + $^{26}$Mg)/$^{24}$Mg with greater certainty than either of the
individual ratios $^{25}$Mg/$^{24}$Mg and $^{26}$Mg/$^{24}$Mg. 
Unfortunately, as our illustrations of short segments of the spectra
suggest, such unblended lines do not exist. Our tests indicate
that the $^{26}$Mg/$^{24}$Mg ratio is more accurately determined from
our spectra than the $^{25}$Mg/$^{24}$Mg ratio.

Isotopic magnesium abundances have been reported previously for
six of the program stars. In general, our results from higher quality
spectra are consistent with  published values. Tomkin \&
Lambert (1980) analysed Gmb\,1830 and $\mu$\,Cas but used only one
of the three features subsequently recommended by McWilliam \& Lambert (1988),
and in the former case included three secondary MgH lines. Our and the
earlier results for Gmb\,1830 are identical:
 $^{24}$Mg:$^{25}$Mg:$^{26}$Mg = 93:4:3. For
$\mu$\,Cas, we find a slightly lower concentration of the heavier isotopes:
85:7:8 now vs 80:10:10 ($\cong$ solar) then. Our use of two additional
recommended MgH lines and higher quality spectra may account for the 
small difference between these results. 

Barbuy et al. (1987) analysed HD\,23439B\footnote{Barbuy et al. refer to
the star observed as HD\,23439B but list the V magnitude of HD\,23439A.
Star B is 0.6 magnitudes fainter than A.}
 and HD\,25329. Our results for HD\,23439A of
84:8:8 are in very good agreement with the ratios 86:7:7 by Barbuy et al.
The 
result is similar for HD\,25329:  85:8:8 from Table 1 
and 90:5:5 from Barbuy et al. Comparison
of the observed spectrum with that of
Gmb\,1830 clearly shows the greater presence of $^{25}$Mg and $^{26}$Mg
in HD\,25329. Similarly, Figure 6 shows that HD\,23439A has markedly
greater concentration of $^{25}$Mg and $^{26}$Mg than Gmb\,1830.
Barbuy et al. (1987) show the spectrum of HD\,25329 around 5135\AA.
The higher resolving power of our spectrum is clearly seen by
comparing the central depths of the lines which are deeper
in our spectrum (e.g., the MgH line at 5134.6\AA\ has a central depth of
0.46 against 0.58 in Barbuy et al.'s spectrum), and in the clearer
presence of the 
inflections and partially resolved lines due to $^{25}$MgH and $^{26}$MgH
in our spectrum. In addition, the $^{26}$MgH line in our spectrum
 is well resolved in the
5140.2\AA\ feature.

Tau Ceti and HR\,1299 were analysed by McWilliam \& Lambert (1988).
For $\tau$ Ceti, our ratios are 75:15:9 and the former estimates 
were 83:7:10. Barbuy (1985) obtained 84:7:7. For HR\,1299, our result
72:18:10 compares with the previous result 80:10:10. In both cases,
 the disagreement concerns 
 the
$^{25}$Mg/$^{24}$Mg ratio. Inspection of a set of syntheses for these and
other stars with more $^{25}$Mg than $^{26}$Mg
 indicates that an isotopic mixture with nearly equal
parts $^{25}$Mg and $^{26}$Mg does not provide a satisfactory fit to the
observed spectra. We suppose that the higher resolution of our spectra
enables us to derive the $^{25}$Mg abundance where others may have
assumed that $^{25}$Mg and $^{26}$Mg were of similar abundance.

\bigskip
\section{Results and Discussion}

\subsection{Chemical Evolution of $^{25}$Mg and $^{26}$Mg}

Our isotopic ratios may be compared with predictions based on the
assumption that the
magnesium isotopes are exclusively
 a product of hydrostatic burning in massive stars
and are ejected by the supernova explosion. We comment later on a possible
contribution from intermediate-mass AGB stars that synthesise $^{25}$Mg
and $^{26}$Mg in the course of operating the $s$-process by the
neutron source $^{22}$Ne($\alpha,n)^{25}$Mg.  
Extensive calculations of nucleosynthesis by massive stars (Woosley \&
Weaver 1995) may be used to illustrate
the  predicted metallicity dependence
of the isotopic Mg ratio. We take as a representative
model of  a 30$M_{\odot}$ star from their series B. Woosley \&
Weaver characterize the composition of the ejecta in two ways, as mass
 in $M_{\odot}$, and as a production factor which is the mass fraction
of a nuclide contained in the ejecta relative to mass fraction
in the adopted standard solar mix. If the production factors predicted
for a collection of different nuclides are numerically equal, these
nuclides are produced in standard proportions. For our present purpose,
we compare the ratio (here, $p_{25}$)
 of the production factors of $^{25}$Mg and $^{24}$Mg
for the 30B models of different initial composition. The ratios for
$^{26}$Mg and $^{24}$Mg behave very similarly. Obviously, in the case that
$p_{25}$ = 1, this isotopic ratio in the ejecta is the standard value.
The predicted values are 0.78, 0.086, 0.025, 0.024, and 0.056 for models
having the initial mass fraction $Z$/$Z_{\odot}$ = 1, 0.1, 0.01, 10$^{-4}$, and
0.0. The value of $p_{25}$ at a given $Z$ is weakly dependent on
a model's mass.
The initial decline
of $p_{25}$ from $Z = Z_{\odot}$ to $Z = 0.1Z_{\odot}$ is presumably
the result of $^{25}$Mg production via $^{22}$Ne($\alpha,n)^{25}$Mg and the
decline in the abundance of the seed $^{22}$Ne. A shallowing of the
decline for $Z < 0.01Z_{\odot}$ likely reflects a primary production
of the $n$-rich isotopes. 
 Since $p_{25}$ is weakly
dependent on stellar mass, the predicted isotopic ratios are rather
insensitive to the assumed form of the initial mass function,
and evolution of the $^{25}$Mg/$^{24}$Mg
 and $^{26}$Mg/$^{24}$Mg in the Galaxy may
be qualitatively predicted from this discussion.

 One expects a ratio of
$^i$Mg/$^{24}$Mg of a few per cent from very low metallicities  until
$Z \simeq 0.01Z_{\odot}$ and then an increase that steepens as
$Z = Z_{\odot}$ is reached. Timmes et al.'s (1995)
comprehensive theoretical study of the chemical evolution of the
solar neighborhood includes quantitative  predictions of the isotopic ratios
$^{25}$Mg/$^{24}$Mg and $^{26}$Mg/$^{24}$Mg. 
Predicted trends of $^{25}$Mg/$^{24}$Mg and $^{26}$Mg/$^{24}$Mg vs [Fe/H]
are almost identical with $^{26}$Mg/$^{24}$Mg about 6\% greater at a
given [Fe/H].
Timmes et al. considered the ejecta of Type II
  and Type Ia supernovae. Both types of 
supernovae return iron
 to the interstellar medium but only the Type II supernovae 
return magnesium in appreciable quantities.
These predictions (Figures 7 and 8) for the isotopic Mg abundances
appear to be the only extant predictions that
incorporate predicted compositions of supernovae ejecta in a model of the
solar neighborhood. 

 Contributions from   intermediate  mass
stars were not considered by Timmes et al.
 These stars  may synthesise $^{25}$Mg and
$^{26}$Mg when, as AGB stars,
they experience the $s$-process.  These products get into  
the interstellar medium when an AGB star sheds its envelope.
Constraints on contributions from AGB stars should be obtained from
observations of other abundances that they may contribute, principally N
and the $s$-process elements.  It may then be important to the
story of the Mg isotopic abundances that
Timmes et al.'s standard predictions failed
to fit the observation that [N/Fe] $\simeq 0$ over the range 
-2.5 $<$ [Fe/H] $<$ 0 for which N abundances are presently known. These
predictions show [N/Fe] declining steeply for [Fe/H] $\leq$ -1.5. 
Possibly, the nitrogen deficiency of the galactic chemical evolution
model may be made up from intermediate-mass stars which may manufacture
large amounts of N by CN-cycle H-burning of C dredged-up from the
He-shell, particularly in a terminal evolutionary phase when the
star develops a `hot-bottomed-convective-envelope'.

 Pagel \&
Tautvai\v{s}ien\.{e} (1997) propose a model to account for the observed
abundances of $r$ and
$s$-process  species that includes an $s$-process  contribution from AGB stars.
Norris (1999) suggests that these stars may also account for the early
evolution of the N abundances.
 Massive stars are presumed responsible for the
$r$-process. Pagel \& Tautvai\v{s}ien\.{e} suppose the 
 $s$-process contributions  to be
delayed with respect to the
$r$-process. Delays of about 37 Myr and 2.7 Gyr were found from a fit to the
observed abundances, that is progenitor masses of  about
8$M_{\odot}$ and 1.5$M_{\odot}$ respectively. The effects on galactic
abundances of the ejecta from the 1.5$M_{\odot}$ stars occurs only for
[Fe/H] $\geq$ -0.7. These low-mass stars most probably are ineffective
contributors of the Mg isotopes and certainly not competititve with 
massive stars. 
 From their first contributions at [Fe/H] $\sim$ -2.4 to
[Fe/H] = -0.7, the 8$M_{\odot}$ stars control $s$-process abundances.

Copious amounts of $^{25}$Mg and
$^{26}$Mg are expected from the 8$M_{\odot}$ stars that should run the
$s$-process with the $^{22}$Ne($\alpha,n)^{25}$Mg neutron source. Yields of
$s$-process species should scale approximately with a star's initial
metallicity, i.e., the $s$-process is, in the language of the subject, a
secondary not a primary process. However, Pagel \& Tautvai\v{s}ien\.{e}'s
fit to the observed abundances
requires  primary production of $s$-process species by the
8$M_{\odot}$ (and 1.5$M_{\odot}$) stars. This, as they noted,
would suggest that an alternative neutron source, 
probably the $^{13}$C($\alpha,n)^{16}$O reaction,
operated with very little contribution from the $^{22}$Ne source. The 
$^{13}$C source is expected to be dominant in any case in the 1.5$M_{\odot}$
stars. Although detailed calculations should be made, it is likely
that operation of the $^{13}$C source does not lead to major enrichments
in the stellar envelope (and subsequently in the ejecta) of $^{25}$Mg and
$^{26}$Mg relative to $^{24}$Mg. If true, the expectation is that
the massive stars control galactic chemical evolution of the Mg isotopes
despite their probable failure to account for the observed N
abundances.

 Thanks to the
larger isotopic wavelength shift, the $^{26}$MgH lines
are less blended with the strong $^{24}$MgH parent lines than the
$^{25}$MgH lines. Then, the $^{26}$Mg abundance is likely more
reliably determined than the $^{25}$Mg abundance. We begin by
comparing the observed and predicted $^{26}$Mg/$^{24}$Mg ratios.
Our sample contained  a minority with known abundance anomalies
 (e.g., the N-rich subdwarf HD\,25329) but the majority were known to
have normal abundances for their metallicity. It was anticipated that
some of the minority might also have anomalous isotopic ratios and so it
proved. What was not anticipated was that stars considered or assumed to
be of normal composition for their metallicity would possess anomalous
isotopic ratios. 

In the  comparison made in Figure 7 between predicted and measured
$^{26}$Mg/$^{24}$Mg ratios, we distinguish normal and peculiar stars
where the latter group comprises  stars with anomalous elemental
abundances for light and/or heavy elements but not all such peculiar stars
have odd isotopic ratios.
Figure 7
shows that
the measured isotopic ratio for the normal stars
and the predictions are in general agreement: the ratio
$^{26}$Mg/$^{24}$Mg declines with decreasing [Fe/H] in line with the
predicted trend. It is apparent too that there is a  scatter across the
sample of normal and peculiar stars showing that
the isotopic ratio at a given [Fe/H]  exceeds the measurement
errors of the isotopic ratio and [Fe/H]. That the scatter is in part intrinsic
is well shown by Figure  6 where spectra of Gmb\,1830  and HD\,23439A are
plotted. It is seen that the strengths of the $^{24}$MgH-dominated cores of
the two identified recommended
MgH features (also, the MgH feature at 5138.35\AA) are very similar but
the red wings to which $^{25}$MgH and $^{26}$MgH contribute are
pronounced for HD\,23439A but not for Gmb\,1830. Since the atomic line
at 5137.4\AA\ is symmetrical in both spectra, the stronger asymmetry must
be due to the heavier Mg isotopes and to a greater isotopic ratio for
these isotopes in HD\,23439A. Peculiar stars like HD\,23439A 
are discussed
following remarks on the general trend of $^{26}$Mg/$^{24}$Mg and
$^{25}$Mg/$^{24}$Mg with [Fe/H]. The sample of normal stars may be too
small and the errors of measurement of such a size that the evidence
for intrinsic scatter is not quite so striking. 

 There is  a hint that the observed ratio $^{26}$Mg/$^{24}$Mg is
consistently greater than predicted.
% especially at the  
%low metallicity limit of our sample where the $^{26}$Mg/$^{24}$Mg
%ratio of the three most metal-poor stars   is about
%a factor of 4 less than the solar system's ratio but  greater
%than the prediction by about a factor of 2. At the high metallicity end,
%the prediction exceeds the solar ratio.
 Timmes et al. note that their
prediction of the iron yield from Type II supernovae depends on the
mass cut between the collapsed object (neutron star or black hole) and the
ejecta. Since yields of lighter elements, including magnesium, are
unaffected by the
positioning of the mass cut,  the effect of a change in the
mass cut is to translate the predicted
$^{26}$Mg/$^{24}$Mg vs [Fe/H] along the [Fe/H] axis; 
 for example, a slightly shallower mass cut independent of a star's initial
composition
 reduces the [Fe/H] that
corresponds to a given isotopic ratio. A change of the mass cut with [Fe/H]
would, of course, change the shape of the predicted curve.
There is a second way to affect the predicted relation.
Iron is contributed by both Type II and Type Ia supernovae.
 The  Type Ia
supernovae
 provide iron but very little magnesium and, hence, 
if their  contribution relative to that of the SN II is altered, the
predicted curve in Figure 7 is translated along the [Fe/H] axis. In addition,
the contribution from Type Ia supernovae
 is delayed relative to the almost instantaneous contribution from the Type II
supernovae.
Therefore,  a change in the adopted
lifetime of the Type Ia  precursors changes the shape of the predicted
curve around the metallicity at which
the Type Ia supernovae begin to contribute to [Fe/H], and displaces the curve
over the [Fe/H] range for which the Type Ia and II supernovae are contributing
in tandem; for example, if the lifetime for the precursors is longer than
assumed by Timmes et al., the predicted curve is translated to lower [Fe/H].
But, as noted above,
Timmes et al.'s choices for the various quantities that affect
the translation and shape of the predicted $^{26}$Mg/$^{24}$Mg relation
result in a relation that fits our observations of $^{26}$Mg/$^{25}$Mg ratios
remarkably well. If the Mg isotopes are contributed partly by AGB stars,
an intrinsic scatter in the isotopic ratio at a given [Fe/H] would not
be surprising in light of the evidence that there is an
intrinsic scatter in the ratios [$s$/Fe] at a given [Fe/H] (Edvardsson
et al. 1993).

Measurements of the $^{25}$Mg/$^{24}$Mg are somewhat less precise because
the $^{25}$MgH line is more severely blended with the $^{24}$MgH line
 than is the
$^{26}$MgH line. Figure 8   
shows that $^{25}$Mg/$^{24}$Mg mimics the decline of the  $^{26}$Mg/$^{24}$Mg
ratio with decreasing [Fe/H]. Two differences are noticeable: the scatter
of the measured $^{25}$Mg/$^{24}$Mg ratios about a mean relation is
larger than for the $^{26}$Mg/$^{24}$Mg ratios, and the
mean relation is decidedly offset from the predicted curve.
There is a clear tendency for the measured $^{25}$Mg/$^{24}$Mg 
ratios to be displaced by either about 0.3 dex in [Fe/H] or 0.03 in 
$^{25}$Mg/$^{24}$Mg. As detailed above, simple adjustments to the predictions
may be envisaged to bring observation and theory into agreement: alter
the mass cut in Type II supernovae or change the ratio of Type II to Type Ia
supernovae that contributed to the chemical evolution of the disk.
 The displacement may be slightly larger than
for $^{26}$Mg/$^{24}$Mg. If true, this would require additionally a change in
the relative yields of $^{25}$Mg and $^{26}$Mg from Type II supernovae.
 It would be of interest in this connection to have an assessment of
the sensitivity of the yields to the uncertainties in the key nuclear
reaction rates influencing synthesis of the Mg isotopes. A comparison of yields
from massive stars (Hoffman et al. 1999) suggests that these uncertainties are
certainly not an insignificant contributor to predicted evolution of the Mg
isotopes.

Timmes et al. checked their predictions against measurements
drawn from the literature of the
$^{25}$Mg/$^{24}$Mg ratio in dwarfs and subgiants.
 Timmes et al. aver that
``for [Fe/H] $\geq$ -1.0, the calculations agree fairly well
with the magnesium isotope ratios found in disk dwarfs''.
 Inspection of the
comparison (their Figure 16) suggests, however,  that the scatter of the
collated  observations is so large that ``fairly  well'' must be an elastic
qualifier.
 Our
present results would likely cause Timmes et al. to reiterate their
remark but with the qualifier `fairly' expunged!

 For low  metallicities, [Fe/H] $\leq -1.0$, Timmes et al.
selected three additional data points: Tomkin \& Lambert's (1980)
original result for Gmb\,1830 which fell near the predicted curve
 and two results from 
 Barbuy (1985) that do not at all fit the predicted
curve: HD\,24616 at [Fe/H] = -1.5 had slightly sub-solar abundances
of $^{25}$Mg and $^{26}$Mg and HD\,188510 at [Fe/H] = -1.8 had
solar isotopic ratios. Inspection of the literature shows that the
iron abundance of HD\,24616 is higher than adopted by Barbuy. When
[Fe/H] = -0.8 (Fran\c{c}ois 1988; Pilachowski, Sneden \& Booth  1993) 
is adopted, Barbuy's isotopic ratios fit the predicted curves.
HD\,188510 is not so simply brought into
conformity; the adopted metallicity is confirmed by recent spectroscopic
analyses, e.g., Beers et al. (1999) select [Fe/H] = -1.53 from a literature
survey. Our observations of this star at a resolving power of 200,000 show
very weak MgH features: the 5138.7\AA\ feature is just 5\% deep in good
agreement with a prediction based on Barbuy's adopted atmospheric parameters.
 If  isotopic
ratios are solar, the $^{25}$MgH and $^{26}$MgH lines would be
at most about 0.5\% deep. Our exploratory spectrum had a S/N ratio of about
60 so that no information except high upper limits can be given for the
ratios. Barbuy's spectrum was at a resolving power of 80,000 - 100,000 
and the quality is given as `bad' which would seem to imply a S/N of 100
or less at which the quoted isotopic ratios are not determinable unless the
MgH were very much stronger than at the time of our observation.

Timmes et al.  overlooked a few measurements:
Lambert \& McWilliam (1986) set an upper limit of 3\% to the
$^{25}$Mg/$^{24}$Mg  and $^{26}$Mg/$^{24}$Mg ratios for the subgiant
$\nu$ Indi with [Fe/H] = -1.5; Barbuy et al. (1987) reported 
isotopic ratios of 6 to 8\%
for three additional stars with [Fe/H] in the range -1.1 to -1.6. The upper
limits are consistent with the predictions and our measurements. Barbuy et al.'s
measured ratios exceed the predictions. 

Observers' views of the evolution of the Mg isotopic ratios
have evolved. Tomkin \& Lambert (1980) discovered the low abundance
of the neutron-rich isotopes in Gmb\,1830, a result confirmed here, and
noted that it was consistent with synthesis of the Mg isotopes by
massive stars.
In contrast, Barbuy's (1985) pioneering survey of the Mg isotopes in 24 stars
led her to note that ``values [of the isotopic ratios] close to the
solar ratios $^{24}$Mg:$^{25}$Mg:$^{26}$Mg = 79:10:11 are
generally found'' for the range -1.5 $<$ [Fe/H] $<$0. 
 Barbuy et al. (1987), who analysed
three halo dwarfs  and two super-metal-rich dwarfs, combined their
results with those in the literature to suggest a trend for the 
isotopic ratios: $^i$Mg/$^{24}$Mg where $i$ = 25 or 26
 has the solar ratio for [Fe/H] $>$ -0.5,
and is lower by 0.4 dex for halo stars with [Fe/H] $<$ -1.0 with a 
smooth transition between the solar and halo ratios between [Fe/H] of -1.0
and -0.5. Such a relation provides an adequate fit to our
results for the $^{25}$Mg/$^{24}$Mg ratio but our  results
for the $^{26}$Mg/$^{24}$Mg ratio, which are more accurate than those
for $^{25}$Mg/$^{24}$Mg clearly indicate a continuous decline in
the ratio with decreasing [Fe/H] on the plausible assumption that
Gmb\,1830 and $\nu$\,Ind rather than HD\,25329 and its
peculiar cohorts are taken to define the decline. In light of other
abundance peculiarities displayed by HD\,25329 and several other peculiar
stars, the galactic evolution of the Mg isotopes is most probably
better defined by normal stars like Gmb\,1830.

\subsection{Peculiar Stars}

Our intent was that the label  `peculiar' should be attached only to
stars in our sample known to have anomalous abundances of one or more
elements. By this definition, the peculiar stars are  
 HR\,1299,  HD\,23439A and B, HD\,25329,  and  
 HD\,134439 and HD\,134440. An additional
star -- HD\,108564 -- is termed peculiar
on the basis of above average relative abundances of $^{25}$Mg and $^{26}$Mg. 
%HD\,108564 may possibly have other abundance anomalies -- see below.

HD\,108564 is notable for the fact that  the spectroscopic
estimate of [Fe/H] is appreciably lower than the photometric estimate.
 Our syntheses
give [Fe/H] = -0.9 from a few strong lines. Ryan \& Norris (1991) estimate
[Fe/H] = -1.12 from the U - B ultraviolet excess, an estimate oddly
inconsistent with the Str\"{o}mgren photometry but consistent with 
the spectroscopic estimate.
 The star is listed as a CH subdwarf   by
Bartkevi\v{c}ius (1996) but it is then surprising that 
Tomkin \& Lambert
(1999) who specifically investigated  heavy elements did not report
their overabundance.
Twarog \& Anthony-Twarog (1995) report photometric measurements of the
Ca\,{\sc ii} H and K lines. These show the star to be quite unusual
in that the $hk$-index is stronger than is measured for [Fe/H] $\sim 0$
stars; metal-poor stars show the expected weaker $hk$-index.
 This anomalous
$hk$-index may be indicative of unusual molecular line blanketing. 
 Further spectroscopic examination
is warranted. We shall assume that [Fe/H] = -1.18 is the star's metallicity.
HD\,108564 is greatly overabundant in $^{25}$Mg and $^{26}$Mg for this
metallicity. The isotopic ratios are normal for the photometric metallicity.

Binary members HD\,23439A and B were shown by Tomkin \& Lambert (1999)
to be enriched in heavy elements relative to other stars of the same
metallicity. Enrichments of the measured elements were 0.3 to 0.5 dex for
Y, Zr, Ba, and Nd. The pair are mild CH (dwarf)  stars whose
abundance anomalies are likely to be primordial rather than, as now
generally assumed for CH  and Ba stars, a product
of mass transfer across a binary system; the binary is a wide binary
and, in addition, HD\,23439B is itself a spectroscopic binary. Both 
stars are enriched in $^{26}$Mg
 relative to the mean trend, and HD\,23439A but not
B is enriched in $^{25}$Mg 
 relative to other stars of the same metallicity. The $^{26}$Mg/$^{24}$Mg
ratios of A and B differ by only 2.5\%, which is approximately the
error of measurement. The difference for $^{25}$Mg/$^{24}$Mg is about three
times larger: 8\% (
17\% for A vs 9.5\% for B). To highlight this difference in isotopic ratios,
we provide in  Figures 9 observed and synthetic spectra covering two of the
three recommended MgH lines. For each star we provide the synthetic spectrum
(continuous line) computed for that star's derived isotopic ratios and the
spectrum (dashed line) computed for its companion's derived ratios.  
In the case of HD\,23439A, the ratios for HD\,23439B lead to a slightly
weaker red wing of the MgH lines. Unless the restriction that broadening
mechanisms be symmetric is relaxed, it is probable that A's isotopic
mix is not equal to that of B. In the case  of HD\,23439B, the MgH lines
are stronger and the differences between the synthetic spectra for the
A and B ratios are larger. Nonetheless, it is a concern that the
principal difference in the isotopic ratios concerns the $^{25}$Mg/$^{24}$Mg
ratio  derived from the $^{25}$MgH lines that are more blended with the
$^{24}$MgH line than the $^{26}$MgH line. Since it would be remarkable if the
stars did have different isotopic ratios, we intend to obtain spectra of even
higher quality and pay especial attention to the broadening of MgH and
atomic lines including a determination of the microturbulence.

These anomalous enrichments of the heavier Mg isotopes may be associated
with their $s$-process enrichment, and both may be 
identified with an abnormal
contamination of the stars' natal cloud with  ejecta
from intermediate-mass AGB stars.  Thermal pulses (He-shell flashes) in
stars of initial mass of between about 4 and 10$M_\odot$ are predicted to
release neutrons primarily from the reaction $^{22}$Ne$(\alpha,n)^{25}$Mg. 
These neutrons run an $s$-process with some captured by $^{25}$Mg to form
$^{26}$Mg. Dredge-up from the He-shell into the H-rich convective
envelope is predicted to enrich the envelope markedly in $^{25}$Mg and $^{26}$Mg
as well as the $s$-process products (see review by Lambert 1991).
If ejecta from intermediate-mass AGB stars were not severely diluted
with interstellar material before new stars were formed, those stars
would be enriched relative to typical stars of the same metallicity in
both $s$-process elements and the heavier Mg isotopes, as is observed.
This discovery would appear to conclude a long search among 
barium stars for evidence of $s$-processing and isotopic Mg enrichments
arising from operation of the $^{22}$Ne neutron source in
intermediate-mass stars (see, for example, Tomkin \& Lambert 1979;
McWilliam \& Lambert 1988; Malaney \& Lambert 1988; Barbuy et al. 1992).
A thorough abundance analysis of HD\,23439A and B would now be of great
interest.

HD\,25329 is a remarkably  nitrogen-rich star belonging to the class of 
relatively rare
N-rich subdwarfs discovered by Bessell \& Norris (1982). A detailed abundance
analysis of HD\,25329 was reported by Beveridge \& Sneden (1994). With respect
to normal stars of the same metallicity, HD\,25329's principal abundance
anomalies are a N enchancement by 0.5 dex (Carbon et al. 1987), and an 
overabundance of $s$-process elements  similar to those shown by HD\,23439A
and B. Sodium and possibly aluminum may also be slightly overabundant.
Beveridge \& Sneden  thought it likely that the abundance anomalies
resulted from contamination of the natal cloud
with material ejected by AGB stars, as discussed above for HD\,23439A and B.
Isotopic Mg ratios of 85:8:8 are similar to those of HD\,23439A and B but the
$^{25}$Mg/$^{24}$Mg and $^{26}$Mg/$^{24}$Mg ratios are a factor of
three greater than  those measured for Gmb\,1830, a normal star of slightly
greater metallicity. This suggests that HD\,25329 was enriched in the two heavier
Mg isotopes as the other abundance anomalies were formed.  Intermediate-mass AGB
stars are expected to eject N-rich material with an enrichment of these Mg
isotopes. Thus, the Mg isotopic data support and extend Beveridge \& Sneden's
proposal.

HR\,1299 is an example of a weak G-band giant, a rare class of giants  with
an atmosphere highly contaminated with products of the H-burning CN-cycle,
i.e., deficient in $^{12}$C but with a low  $^{12}$C/$^{13}$C ratio and  rich in
nitrogen (Cottrell \& Norris 1978;
Sneden et al. 1978; Day 1980; Lambert \& Sawyer 1984).
 Some stars including HR\,1299 are
Li-rich relative to similar and normal red giants
(Lambert \& Sawyer 1984).   Since carbon is underabundant
by a large factor, blending of MgH lines with C$_2$ lines is effectively eliminated
for HR\,1299 and the accuracy of the isotopic Mg ratios enhanced. Our results show
that the ratios are essentially normal; whatever processes create a weak G-band
giant the isotopic Mg ratios are unaffected (at least for HR\,1299) in contrast to
the situation found for the N-rich dwarf HD\,25329.

Local disk and halo stars show, as a function of [Fe/H], well defined 
smooth trends
in relative abundance: for example, magnesium and
other so-called $\alpha$-elements (Si, Ca, and Ti) are overabundant
in stars with [Fe/H] $\lesssim -1$ such that [Mg/Fe] $\simeq 0.4$. Even
peculiar stars, HD\,25329 for example, may show this overabundance.
The common proper motion pair HD\,134439 and HD\,134440 are exceptional
because they are $\alpha$-poor; King (1997) shows that [Mg/Fe], [Si/Fe],
and  [Ca/Fe]  are each about 0.3 dex less than the values found for
the vast majority of halo stars.
King speculated that HD\,134439 and HD\,134440  were accreted by our Galaxy
from a dwarf spheroidal galaxy whose chemical evolution to [Fe/H] = -1.5
was more greatly
influenced by Type Ia supernovae than was the halo of our Galaxy.
A change in the Type II to
Ia supernova rates alters the [Mg/Fe] at a given [Fe/H]. As long as the
magnesium comes primarily from Type II supernovae, the Mg isotopic
abundance are not expected to be different in $\alpha$-poor  stars
and normal stars of the same [Fe/H] provided that massive stars, the
progenitors of Type II supernovae, were
formed roughly steadily over the  life of the dwarf spheroidal galaxy;  
if massive stars were formed largely in an initial starburst, and Type Ia
supernovae subsequently built up the galaxy's iron abundance, the Mg isotopic
ratios would be those expected of massive stars more metal-poor than the
gas from which HD\,1334439 and HD\,134440 formed.
 Our analysis shows that, although the
elemental abundance of Mg is peculiar, the isotopic Mg ratios are
quite unexceptional being identical to those found for Gmb\,1830,
a subdwarf of similar [Fe/H] with normal abundances of the $\alpha$-elements.
This result is consistent with the accretion hypothesis. 
It might be noted that `accretion' is not demanded either 
by our observations or by the reported abundance anomalies. If our
halo were not everywhere thoroughly mixed, one would expect 
pockets of star formation occurring in regions more heavily contaminated
than the average region with ejecta from a Type Ia supernova. Stars from
such regions would be $\alpha$-poor.

\section{Concluding Remarks}

These measurements of the isotopic Mg abundances provide confirmation of the
predicted reduction of the $^{25}$Mg/$^{24}$Mg and $^{26}$Mg/$^{24}$Mg
ratios with decreasing [Fe/H], and show that the Mg isotopes are
primarily a product of nucleosynthesis in massive stars. 
Unfortunately, measurements on normal dwarfs do not extend to metallicities
less than [Fe/H] $\simeq -1.5$. At present, the sample of stars
known with [Fe/H] $<$ -1.5 have effective temperatures (T$_{\rm eff}$ $>$
4700K) too high for their spectra to contain MgH lines of an adequate
strength for successful isotopic measurements at the
expected isotopic ratios of $^i$Mg/$^{24}$Mg of about 5\% or less.
A dedicated search for cooler metal-poor stars is needed to provide
suitable targets with which to extend our measurements.  It is in the
regime [Fe/H] $<$ -1.5 that intermediate-mass AGB stars may
contribute to the galactic chemical evolution of nitrogen and the
$s$-process elements and possibly also to the evolution of the
$^{25}$Mg and $^{26}$Mg isotopes. 

A second and important result of our survey has been the demonstration
that some stars known to have an anomalous or peculiar composition are
also marked by distinctive isotopic Mg abundances. The N-rich
subdwarf HD\,25329 and the CH stars HD23439A and B are enriched in $^{25}$Mg
and $^{26}$Mg relative to normal stars of the same metallicity. This
enrichment is attributed to local enrichment of their natal clouds
with ejecta from intermediate-mass AGB stars. One other star
-  HD\,108564 --  is
rather similarly enriched but not so obviously distinguished by
other abundance anomalies. Not all stars with anomalous compositions have
peculiar isotopic Mg ratios: the weak G-band giant HR\,1299 and the
low-$\alpha$ common proper motion pair HD\,134439 and HD\,134440 have
normal isotopic abundances. This result highlights the fact that
the more detailed the information on the chemical composition the
greater are the constraints that may be placed on explanations for the
abundance anomalies.

We are grateful to Jocelyn Tomkin for acquiring and reducing some of the
spectra. We thank Vincent Woolf for reducing a majority of the spectra,
and Beatriz Barbuy,  and Bruce Twarog for helpful correspondence or
conversations. We are most grateful to John Norris for a thorough
review of a draft of this paper. This research has been supported
in part by the National Science Foundation (grant AST-9618414) and the
Robert A. Welch Foundation of Houston, Texas.

\clearpage

\figcaption[Gay.fig1.ps]{The spectrum of HD\,55458 (G88-14) from 5132\AA\
 to 5141\AA. Locations of the MgH lines are identified below the spectrum.
Many of these lines are present in the stellar spectrum but weak blends render
them unsatisfactory for the determination of the isotopic abundances.
Three lines used to
determine the isotopic ratios are marked by an asterisk. 
 \label{fig1}}

\figcaption[Gay.fig2.ps]{The spectrum of HD\,55458 (G88-14) from 5134\AA\
to 5136\AA\ showing one of three principal lines used in the isotopic
abundance analysis - the line at 5134.6\AA (the top panel).
 The weak feature near 5135.7\AA\ is
partly due to lines of the C$_2$. The lower panel shows the spectrum
from 5138\AA\ to 5140.7\AA\ including the recommended features at
5138.7\AA\ and 5140.2\AA.
\label{fig2}}

\figcaption[Gay.fig3.ps]{The spectrum of HD\,23439A from 5134.0\AA\
to 5136\AA\ and from 5138\AA\ to 5140.5\AA. The observed spectrum is
represented by filled circles. Synthetic spectra are shown for the
isotopic ratios
$^{24}$Mg:$^{25}$Mg:$^{26}$Mg = 100:0:0 (dashed line), 78:13:9 (solid line,
best fit to the recommended features), and 72:16:12 and
83:10:6 (dotted lines).
\label{fig3}}
 
\figcaption[Gay.fig4.ps]{The spectrum of HD\,25329 from 5134.0\AA\
to 5136\AA\ and from 5138\AA\ to 5140.5\AA. The observed spectrum is
represented by filled circles. Synthetic spectra are shown for the
isotopic ratios
$^{24}$Mg:$^{25}$Mg:$^{26}$Mg = 100:0:0 (dashed line), 84:8:8 (solid line,
best fit to the recommended features), and 89:6:5 and 81:10:9
(dotted lines).
\label{fig4}}

\figcaption[Gay.fig5.ps]{The spectrum of 12\,Oph from 5134.0\AA\
to 5136\AA\ and from 5138\AA\ to 5140.5\AA. The observed spectrum is
represented by filled circles. Synthetic spectra are shown for the
isotopic ratios
$^{24}$Mg:$^{25}$Mg:$^{26}$Mg = 100:0:0 (dashed line), 74:12:13 (solid line,
best fit to the recommended features), and 69:15:16 and 80:9:10 (dotted lines).
\label{fig5}}

\figcaption[Gay.fig6.ps]{Observed spectra of Gmb\,1830 and HD\,23439A showing
two of the recommended MgH features. While the depths of the cores of these
features are almost the same for the two stars, the 
red asymmetry to the MgH lines  caused by the $^{25}$MgH and $^{26}$MgH
components is
striking for HD\,23439A but not for Gmb\,1830.
The dashed lines show the best-fitting synthetic spectra with the
isotopic ratios taken from Table\,1\label{fig6.ps}}

\figcaption[Gay.fig7.ps]{Isotopic ratio $^{26}$Mg/$^{24}$Mg versus [Fe/H]. Predicted
evolution of the ratio from Timmes et al. (1995) is shown by the solid line
with the extension to low [Fe/H] represented by the dashed line
based on the discussion in the text in Sec. 4.1. Our
results are represented by filled and open circles where the latter
denote `peculiar' stars discussed in Sec. 4.2.\label{fig7}}

\figcaption[Gay.fig8.ps]{Isotopic ratio $^{25}$Mg/$^{24}$Mg versus [Fe/H].
Predicted evolution of the ratio from Timmes et al. (1995) is shown by the solid
line with the extension to low [Fe/H] represented by the dashed line based on
the discussion in the text in Sec. 4.1. 
 Our results are represented by filled and open circles where the latter
denote the `peculiar' stars discussed in Sec. 4.2.
\label{fig8}}

\figcaption[Gay.fig9.ps]{Observed and synthetic spectra of HD\,23439A and
HD\,23439B 5138.0\AA\ to 5140.5\AA. In each panel the solid line is
computed for the derived isotopic ratios for that star, and the dashed
line for the derived ratios for the other star. The derived ratios (Table\,1)
are
 $^{24}$Mg:$^{25}$Mg:$^{26}$Mg = 78:13:9 for HD\,23439A and
 $^{24}$Mg:$^{25}$Mg:$^{26}$Mg = 84:8:8. for HD\,23439B.\label{fig9}}


\clearpage

\begin{thebibliography}{}

\bibitem [] {} Alonso,  A., Arribas, S., \& Mart\'{\i}nez-Roger C. 1996,
   313, 873
\bibitem [] {} Arnett, W.D., \& Thielemann, F.-K. 1985, ApJ, 295, 589
\bibitem [] {} Barbuy, B. 1985, A\&A, 151, 189
\bibitem [] {} Barbuy, B. 1987, A\&A, 172, 251
\bibitem [] {} Barbuy, B., Jorissen, A., Rossi, S.C.F., \& Arnould, M. 1992,
    A\&A, 262, 216
\bibitem [] {} Barbuy, B., Spite F., \& Spite M. 1987 A\&A, 178, 199
\bibitem [] {} Bartkevi\v{c}ius, A. 1996, Balt A, 5, 217
\bibitem [] {} Beers, T.C., Rossi, S., Norris, J.E., Ryan, S.G., \&
   Shefler, T. 1999, AJ, 117, 981
\bibitem [] {} Bell, R.A., \& Branch, D. 1970, ApL, 5, 203
\bibitem [] {} Bessell, M.S., \& Norris, J. 1982, ApJ, 263, L29
\bibitem [] {} Beveridge, C.R., \& Sneden, C. 1994, AJ, 108, 285
\bibitem [] {} Boesgaard, A.M. 1968, ApJ, 154, 185
%\bibitem [] {} Bonifacio, P., \& Molaro, P. 1997, MNRAS, 285, 847
\bibitem [] {} Carbon, D.F., Barbuy, B., Kraft, R.P., Friel, E.D.,
   \& Suntzeff, N.B. 1987, PASP, 99, 335
\bibitem [] {} Carney, B.W., Latham, D.W., Laird, J.B., \& Aguilar, L.A.
    1994, AJ, 107, 2240 
\bibitem [] {} Clementini, G., Gratton, R.G., Carretta, E., \& Sneden, C.
    1999, MNRAS, 302, 22
\bibitem [] {} Cottrell, P.L., \& Norris, J.E. 1978, ApJ, 221, 893
\bibitem [] {} Day, R.W. 1980, Ph.D. thesis, University of Texas
\bibitem [] {} de Bi\`{e}vre, P., \& Barnes, I.L. 1985,
    Int. J. Mass. Spec. Ion. Proc. 65, 211
\bibitem [] {} Edvardsson, B., Andersen, J., Gustafsson, B., Lambert, D.L.,
    Nissen, P.E., \& Tomkin, J. 1993, A\&A, 275, 101
\bibitem [] {} Fran\c{c}ois, P. 1988, A\&A, 195, 226
\bibitem [] {} Fuhrmann, K. 1998, A\&A, 329, 81
\bibitem[] {} Hoffman, R.D., Woosley, S.E., Weaver, T.A., Rauscher, T., \& 
     Thielemann. F.-K. 1999, ApJ, in press
\bibitem [] {} King, J.R. 1997, AJ, 113, 2302
\bibitem [] {} Kurucz, R.L. 1993, in {\it Peculiar versus Normal Phenomena in
A-type and Related Stars.},  IAU Colloquim 138, ed. M.M. Dworetsky,
 F. Castelli,\& R.
Faraggiana
	(San Francisco: ASP), 87
\bibitem [] {} Lambert, D.L. 1991, in {\it Evolution of Stars - the Photospheric
   Abundance Connection}, ed. G. Michaud \& A.V. Tutukov, (Dordrecht:Kluwer),
   299
\bibitem [] {} Lambert, D.L., \& McWilliam. A. 1986, ApJ, 304, 436
\bibitem [] {} Lambert, D.L. \& Sawyer, S.R. 1984, ApJ, 283, 192
\bibitem [] {} Malaney, R.A., \& Lambert, D.L. 1988, MNRAS, 235, 695
\bibitem [] {} McWilliam, A., \& Lambert, D.L. 1988, MNRAS, 230, 573
\bibitem [] {} Morell, O. 1994, Ph.D. Thesis, Uppsala University
\bibitem [] {} Morell, O., K\"{a}llander, D., \& Butcher, H.R., 1992, A\&A,
   259, 543
\bibitem [] {} Nissen, P.E., H$\o$g, E., \& Schuster, W.J. 1997,
   in Proc. ESA {\it Hipparcos} Symp., ESA SP-402 (Noordwijk: ESA) 225
\bibitem [] {} Norris, J.E. 1999, preprint
\bibitem [] {} Pagel, B.E.J., \& Tautvai\v{s}ien\.{e}, G. 1997, MNRAS, 288, 108
\bibitem [] {} Pilachowski, C.A., Sneden, C, \& Booth, J. 1993, ApJ, 407, 699
\bibitem [] {} Ryan, S.G., \& Norris, J.E. 1991, AJ, 101, 1835
\bibitem [] {} Schuster, W.J., \& Nissen, P.E. 1989, A\&A, 222, 69
\bibitem [] {} Sneden, C. 1973, Ph.D. thesis, University of Texas
\bibitem [] {} Sneden, C., Lambert, D.L., Tomkin, J., \& Peterson, R.
   1978, ApJ, 222, 585
\bibitem [] {} Timmes, F.X., Woosley, S.E., \& Weaver, T.A. 1995, ApJS, 98, 617
\bibitem [] {} Tomkin, J., \& Lambert, D.L. 1979, ApJ, 227, 209
\bibitem [] {} Tomkin, J., \& Lambert, D.L. 1980, ApJ, 235, 925
\bibitem [] {} Tomkin, J., \& Lambert, D.L. 1999, ApJ, submitted
\bibitem [] {} Tull, R.G., MacQueen, P.J., Sneden, C., \& Lambert, D.L. 1995,
   PASP, 107, 251 
\bibitem [] {} Twarog, B.A., \& Anthony-Twarog, B.J. 1995, AJ, 109, 2828
\bibitem [] {} Woosley, S.E., \& Weaver, T.A. 1982, in {\it Essays in Nuclear
Astrophysics}, ed. C.A. Barnes, D.D. Clayton, \& D.N. Schramm,
(Cambridge: Cambridge University Press), 377
\bibitem[]{} Woosley, S.E., \& Weaver, T.A. 1995, ApJS, 101, 181

\end{thebibliography}
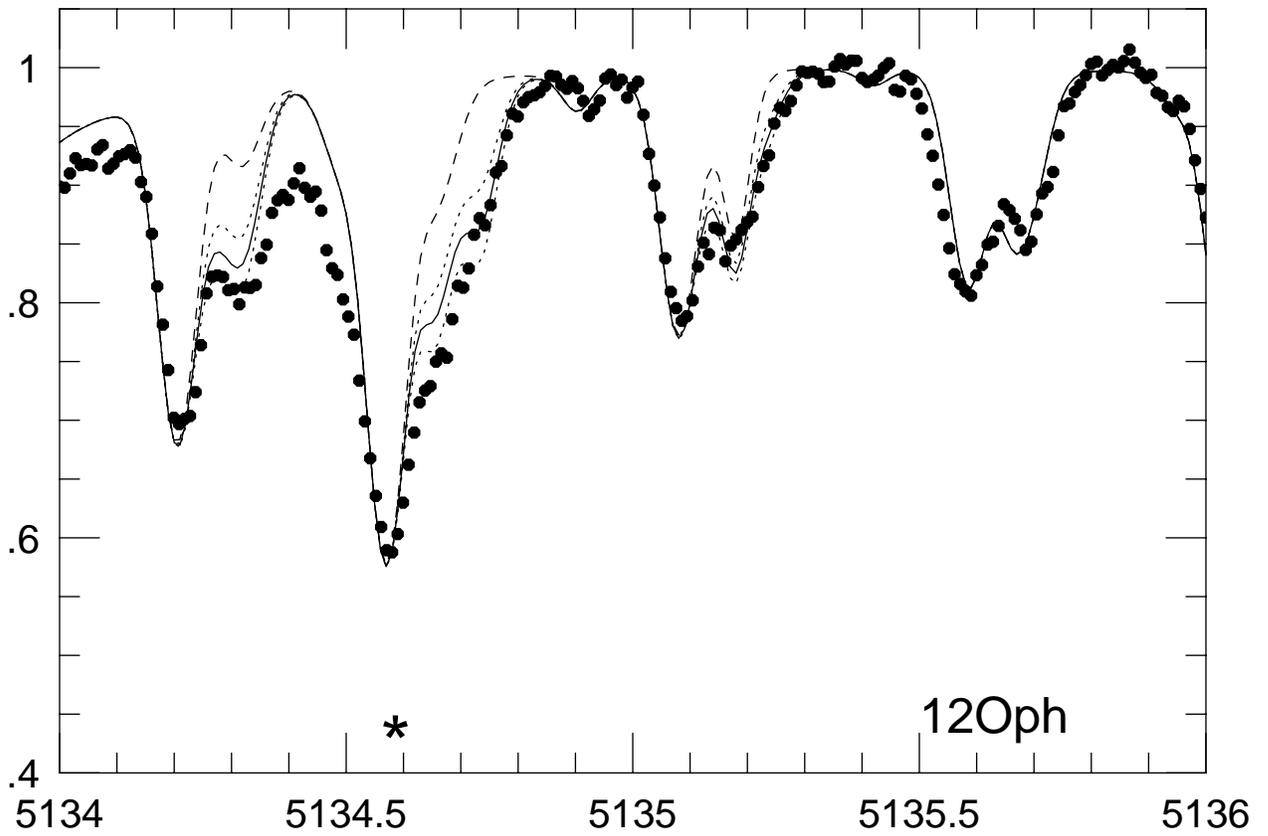
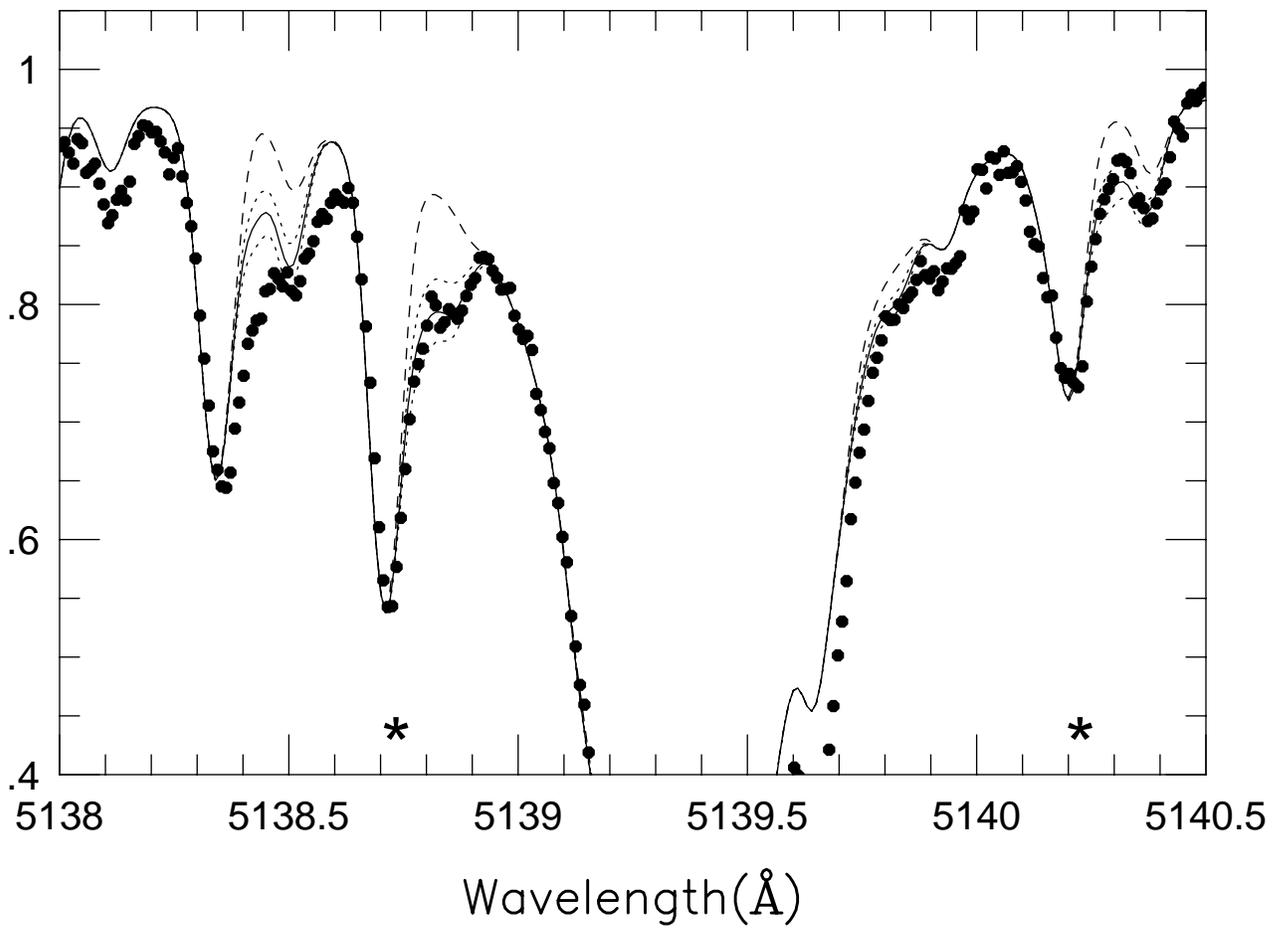
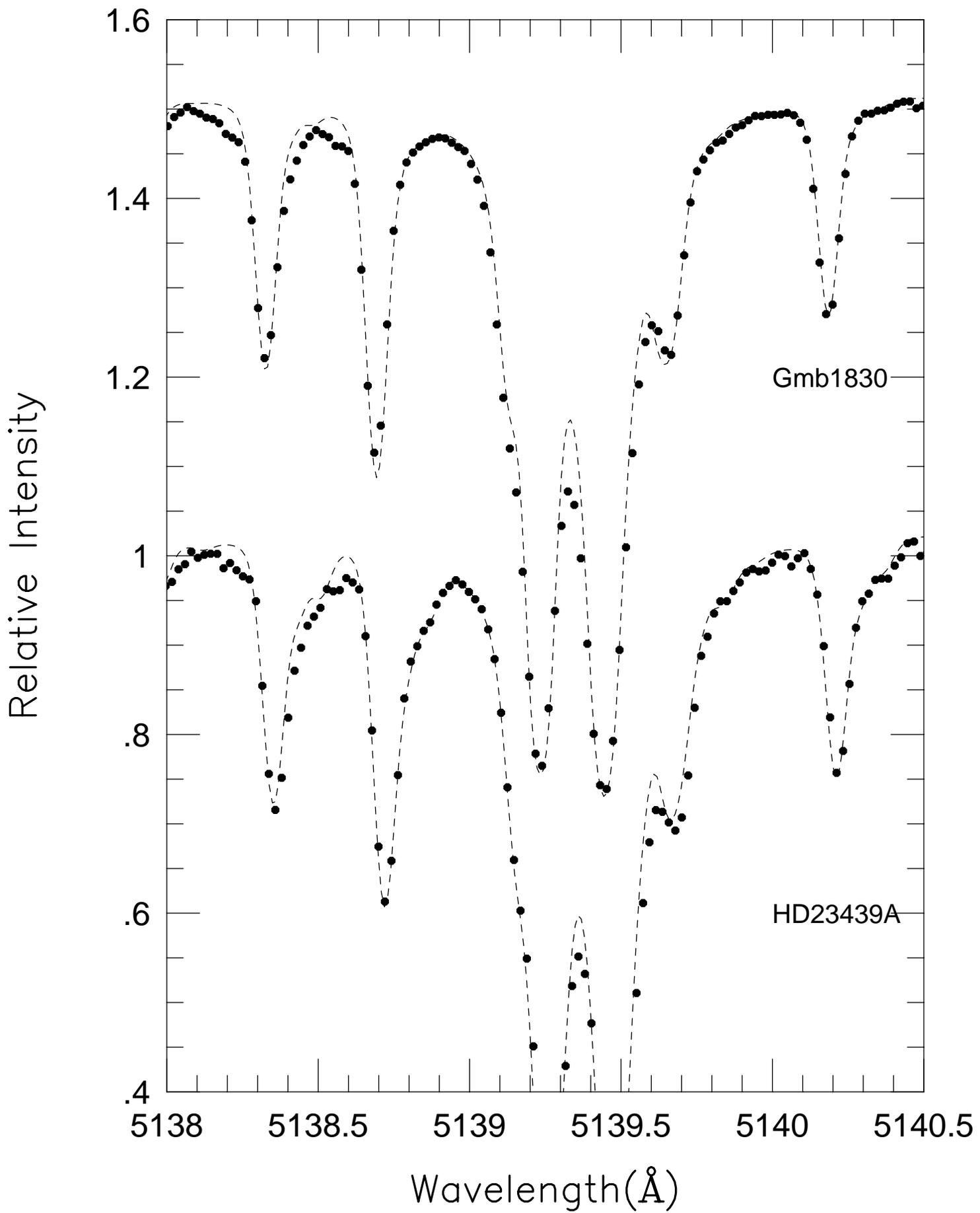
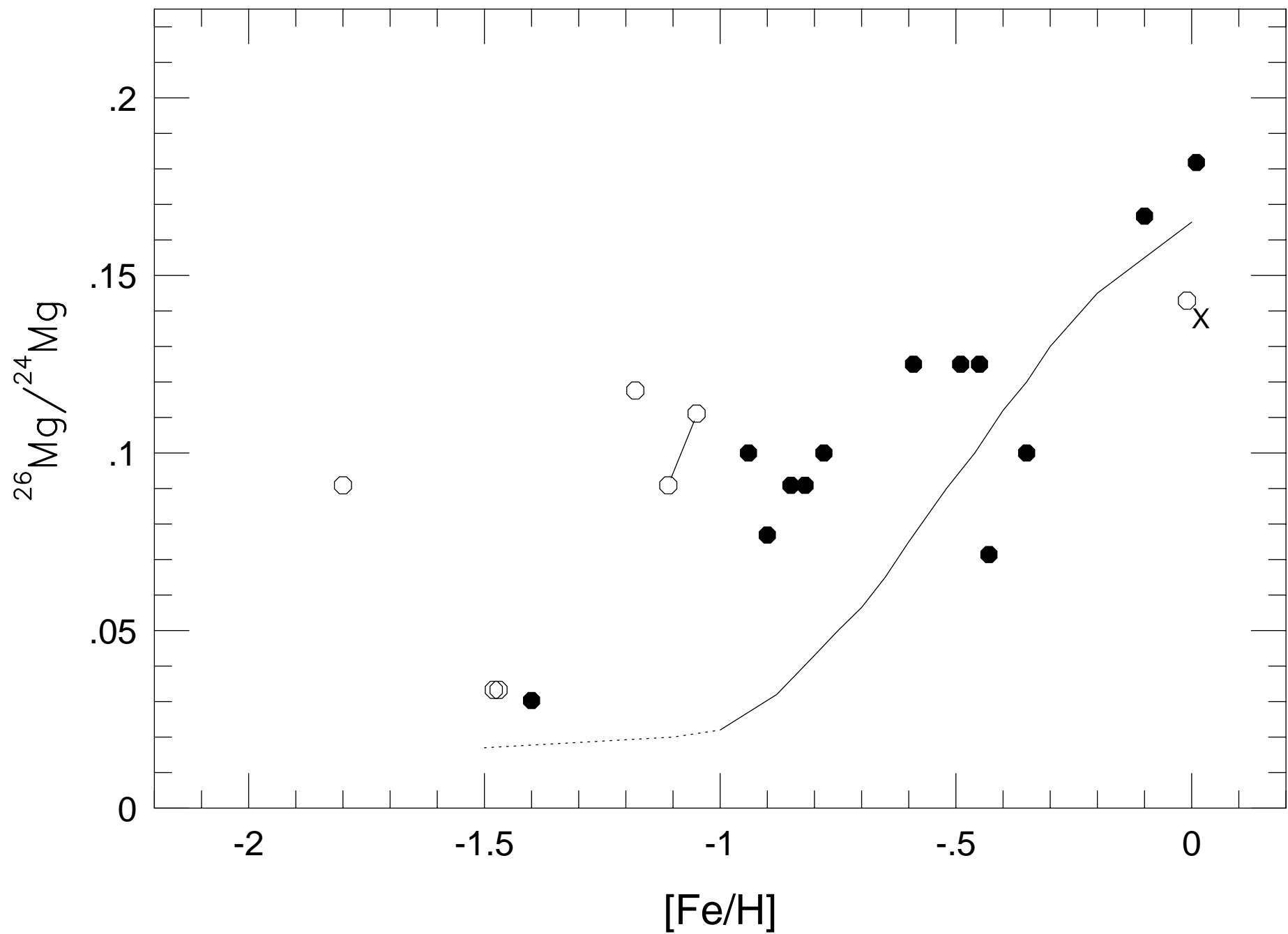
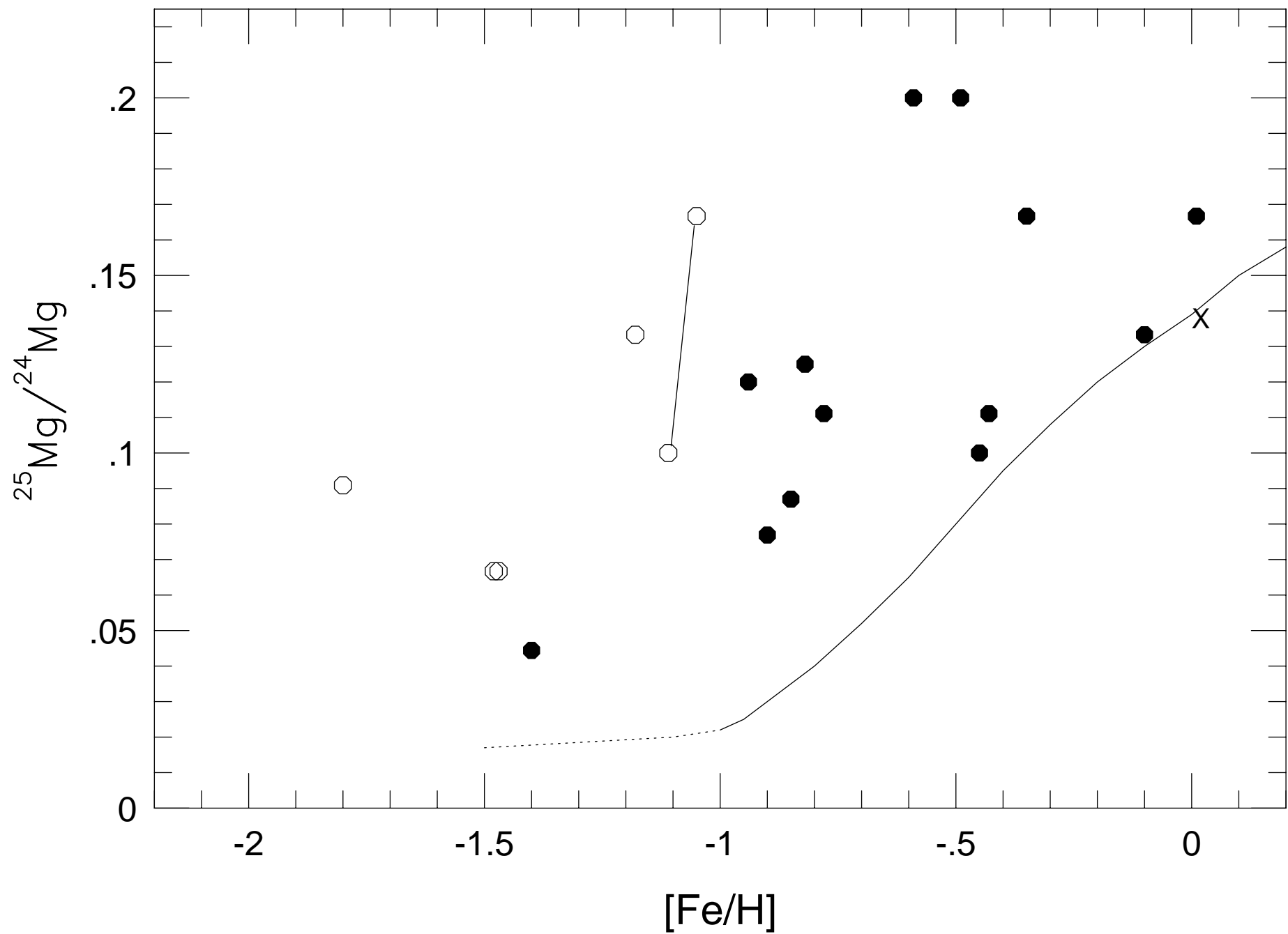
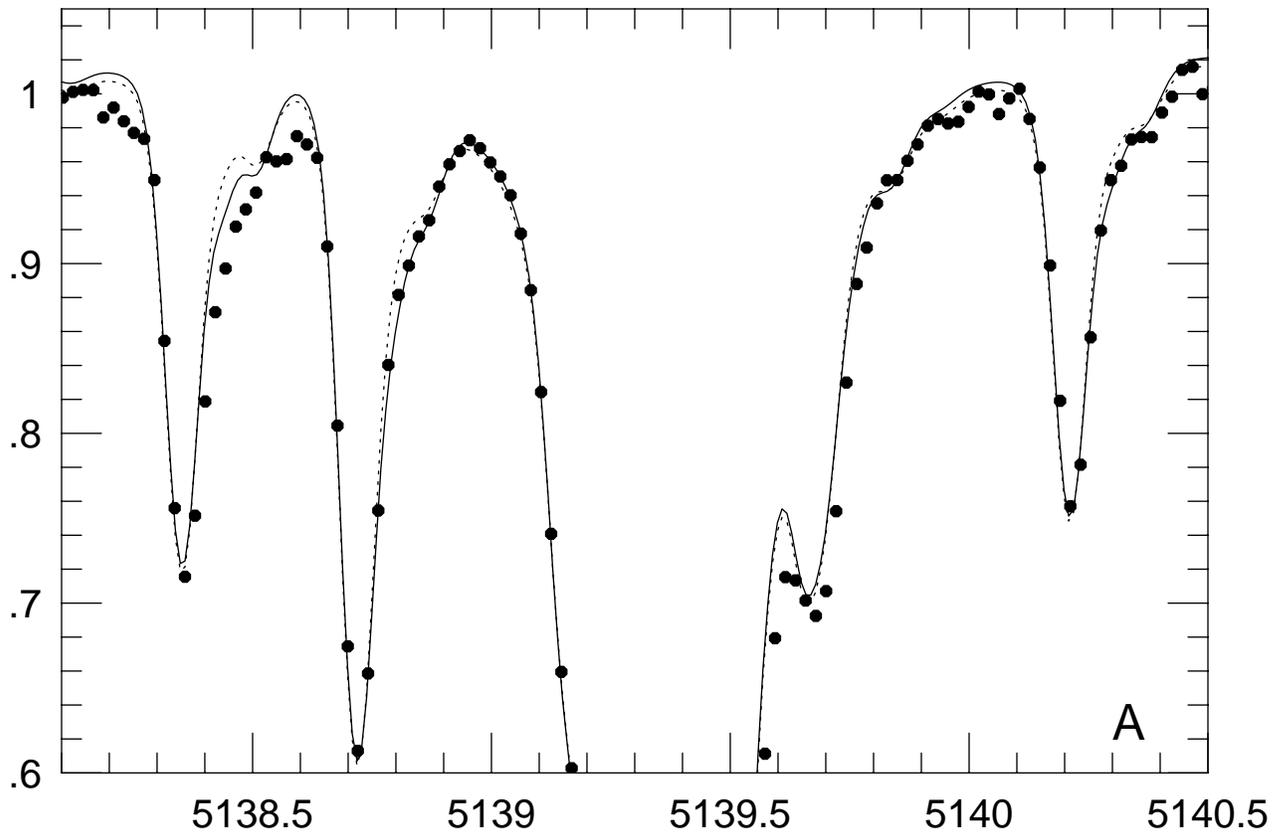
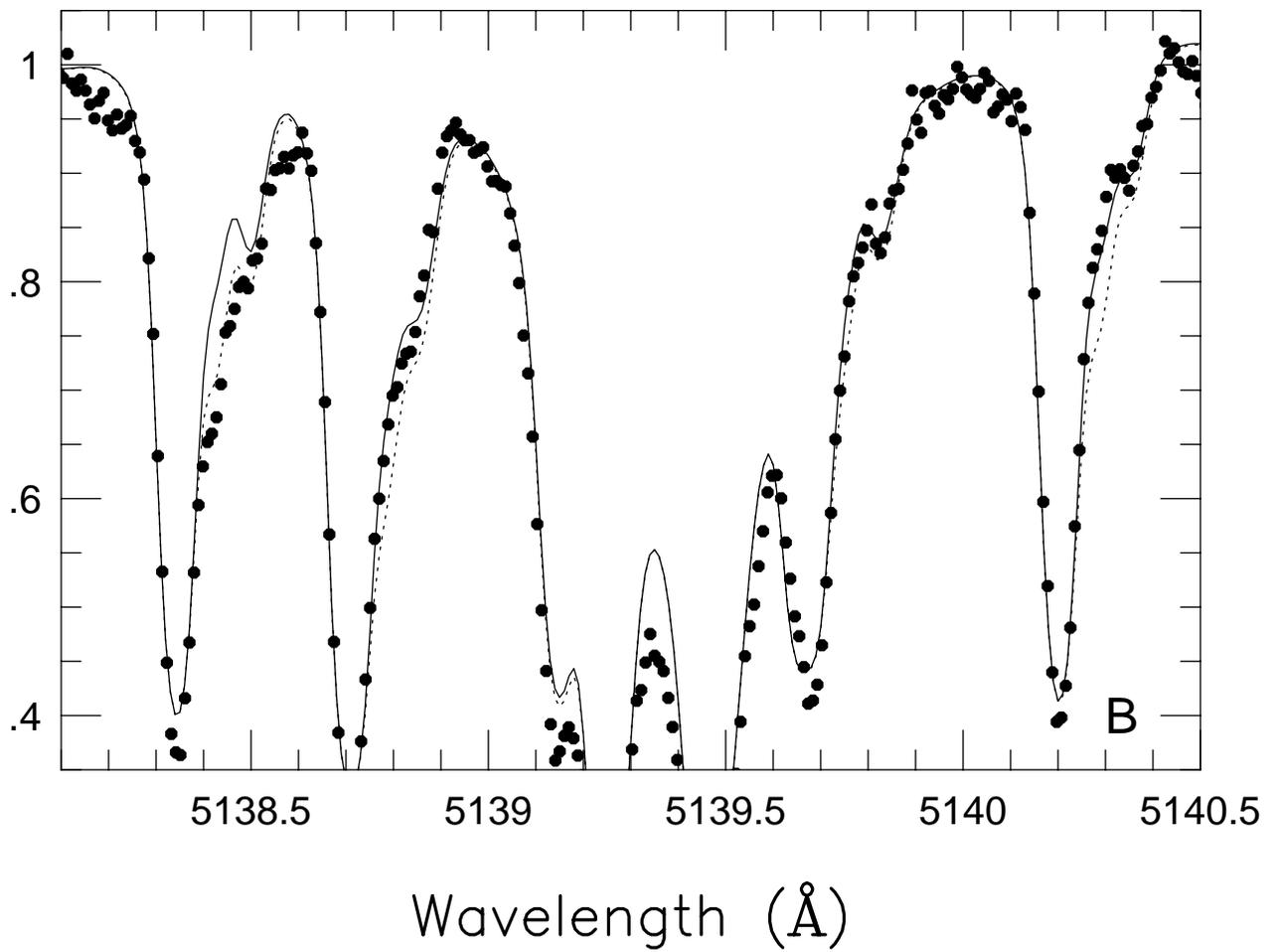
\end{document}